\documentclass[twocolumn,english]{article}
\usepackage[T1]{fontenc}
\usepackage[latin9]{inputenc}
\usepackage{geometry}
\usepackage{amsthm}
\usepackage{amsmath}
\usepackage{amssymb}
\usepackage{graphicx}
\usepackage{cite}

\makeatletter

\usepackage{hyperref}

\makeatother
\usepackage{babel}
\begin{document}

\title{Observation of Pure Spin Transport in a Diamond Spin Wire }

\author{J. Cardellino$^{\dagger,1}$, N. Scozzaro$^{\dagger,1}$, M. R. Herman$^{1}$,
A. Berger$^{1}$, C. Zhang$^{1}$, K.C. Fong$^{1}$, \\C. Jayaprakash$^{1}$,
D.V. Pelekhov$^{1}$, P.C. Hammel$^{1,*}$}

\maketitle
\textbf{Spin transport electronics \textendash{} spintronics \textendash{}
focuses on utilizing electron spin as a state variable for quantum
and classical information processing and storage\cite{wolf2001spintronics}.
Some insulating materials, such as diamond, offer defect centers whose
associated spins are well-isolated from their environment giving them
long coherence times\cite{Hanson18042008,Balasubramanian:2009kx,PhysRevLett.101.047601};
however, spin interactions are important for transport\cite{heinrich2003dynamic},
entanglement\cite{Pfaff:2013fk}, and read-out\cite{grinolds2013nanoscale}.
Here, we report direct measurement of pure spin transport \textendash{}
free of any charge motion \textendash{} within a nanoscale quasi 1D
\textquoteleft{}spin wire\textquoteright{}, and find a spin diffusion
length $\sim700$ nm. We exploit the statistical fluctuations of a
small number of spins\cite{PhysRevLett.91.207604} ($\sqrt{N}<100$
net spins) which are in thermal equilibrium and have no imposed polarization
gradient. The spin transport proceeds by means of magnetic dipole
interactions that induce flip-flop transitions\cite{Bloembergen:SpinDiffusionRelaxation},
a mechanism that can enable highly efficient, even reversible\cite{PhysRevLett.69.2149},
pure spin currents. To further study the dynamics within the spin
wire, we implement a magnetic resonance protocol that improves spatial
resolution and provides nanoscale spectroscopic information which
confirms the observed spin transport. This spectroscopic tool opens
a potential route for spatially encoding spin information in long-lived
nuclear spin states. Our measurements probe intrinsic spin dynamics
at the nanometre scale, providing detailed insight needed for practical
devices which seek to control spin.}

\let\thefootnote\relax\footnote{$^\dag$ These authors contributed equally to this work}
\let\thefootnote\relax\footnote{$^1$ Department of Physics, Ohio State University, Columbus, Ohio 43210, USA.}
\let\thefootnote\relax\footnote{$^*$ Corresponding Author: hammel@physics.osu.edu}

Understanding and controlling spin transport is a central challenge
in spintronics and has been extensively studied in systems where the
spin density is large and treated as a continuum variable\cite{lou2007electrical}.
However, an understanding at the few spin level is needed for miniaturization
and quantum applications. Appropriate and well-controlled coupling
between spins can allow for efficient spin transport, while preserving
long spin lifetimes. For example, two-spin flip-flop transitions \textendash{}
the simultaneous flipping of a pair of anti-aligned neighbouring spins
due to their dipolar interaction \textendash{} conserve magnetization,
and successive flip-flops can result in a pure, diffusive spin transport\cite{Bloembergen:SpinDiffusionRelaxation,abragam},
as reported here. In contrast to non-equilibrium experiments\cite{vinante2011magnetic,eberhardt2007direct}
which measure thermal polarization recovery, we monitor the spin noise\cite{PhysRevLett.91.207604,PhysRevB.84.220405},
which arises from statistical fluctuations. This allows us to observe
the intrinsic thermal equilibrium spin dynamics of the ensemble\cite{PhysRevLett.101.206601}.
\begin{figure*}
\begin{centering}
\includegraphics[width=1.85\columnwidth]{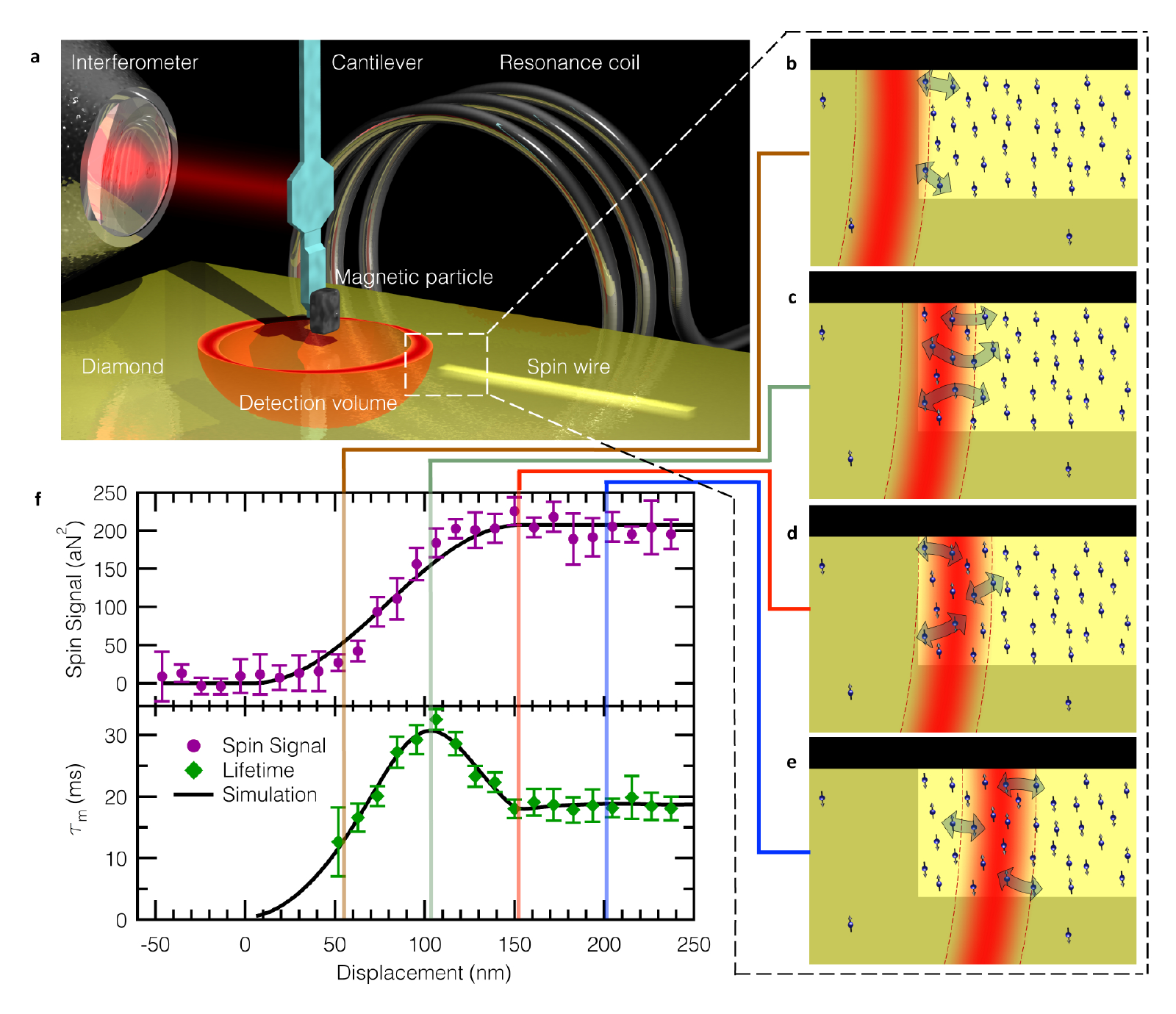}
\par\end{centering}

\caption{\textbf{MRFM spin wire setup, schematic of spin dynamics in wire,
and spin noise measurements}. \textbf{a}, MRFM setup for measuring
spin noise in the 200 nm $\times$ 250 nm $\times$ 4 $\mu$m spin
wire of P1 centres in diamond. The coil excites resonance along a
contour of the magnetic particle\textquoteright{}s field (plus an
external field), called the \textquoteleft{}resonant slice\textquoteright{}.
As the cantilever oscillates, the field due to the magnetic particle
varies and the slice sweeps out the spin detection volume. Spins toward
the centre of the detection volume contribute to the spin signal more
strongly (solid red) than those near the edge. \textbf{b-e}, Schematic
showing the detection volume as it is walked into the spin wire, the
rectangular stripe region. The thick double arrows represent flip-flop-mediated
spin diffusion in and out of the detection volume. \textbf{f}, The
spin signal increases as a function of walk-in distance until the
volume is fully inside the spin wire as in (\textbf{e}). \textbf{g},
The correlation time of the measured spins. At first (as in \textbf{b})
spins are able to quickly diffuse out of the volume, but as the measurement
volume is walked in further (as in \textbf{c}), spins must diffuse
a longer distance to exit the volume, causing the correlation time
to increase. Once the measurement volume is fully inside the stripe
(\textbf{d} and \textbf{e}), spins can diffuse out on both sides,
decreasing the correlation time to roughly half the peak value. }
\end{figure*}

In macroscopic systems containing a large number of particles, Fick\textquoteright{}s
law of diffusion describes the time evolution of particles moving
from regions of high to low concentration, resulting in a smooth evolution
toward equilibrium. By contrast, in few-spin ensembles fluctuations
can lead to transient polarizations much larger than the Boltzmann
(thermal) polarization, and even flow \textquotedblleft{}against\textquotedblright{}
the polarization gradient\cite{seitaridou2007measuring}. Our measurements
focus not on the polarization itself, but on the time auto-correlation
of the polarization in a nanoscale spin detection volume. In our experiments,
these correlations are lost as a consequence of spin transport out
of our measurement volume. We model this process using numerical simulations
that incorporate this inherent randomness of few-spin, statistically-polarized
ensembles, and our model corroborates our measurements. 

In order to study flip-flop-dominated spin dynamics, we fabricated
a nanoscale channel of implanted nitrogen (P1) centres in diamond
at a density of 6 ppm. At this density the strength of the dipolar
coupling between spins is strong, leading to a flip-flop time, $T_{{\rm ff}}\sim10^{-4}$
s, several orders of magnitude shorter than $T_{1}\sim1$ s. The density
outside the spin wire ($0.3$ ppm) is sufficiently small such that
flip-flop transitions out of the wire are very rare, thus confining
the spin transport to the wire. 

We use magnetic resonance force microscopy (MRFM) to probe the spins
within the wire (Fig. 1a). The force detector is an IBM-style ultra-soft
cantilever\cite{chui2003mass} with a SmCo$_{5}$ magnetic particle
glued onto the tip. To resonantly detect spins, we implement the iOSCAR
timing protocol\cite{PhysRevLett.91.207604,r:singlespin}. A superconducting
niobium coil is used to drive the spin resonance at a frequency of
$\omega_{{\rm rf}}=2.18$ GHz. This frequency defines a \textquoteleft{}resonant
slice\textquoteright{} where the total magnetic field (tip field plus
external field) experienced by sample spins is equal to $\frac{\omega_{{\rm rf}}}{\gamma}=778$
G. By driving the cantilever at its self-determined resonant frequency
to an amplitude of $150$ nm, the resonant slice sweeps out a volume
of this width, hereafter termed the detection volume.

We scan the detection volume into the spin wire (Fig. 1b-e) and measure
the force exerted on the cantilever by the selected spins. From this
measurement we can extract two quantities: spin signal (Fig. 1f, top),
and the force correlation time $\tau_{{\rm m}}$ (Fig. 1f, bottom).
The spin signal is obtained from the variance in the time record of
the force signal and is directly related to the number of measured
spins\cite{r:singlespin}. $\tau_{{\rm m}}$ describes the characteristic
time for the net moment of the detected spins to decorrelate. This
can be viewed as the time needed to transport spin out of the detection
volume via a random-walk diffusion processes, with each step taking
an average time$T_{{\rm ff}}$. 
\begin{figure*}
\begin{centering}
\includegraphics[width=2\columnwidth]{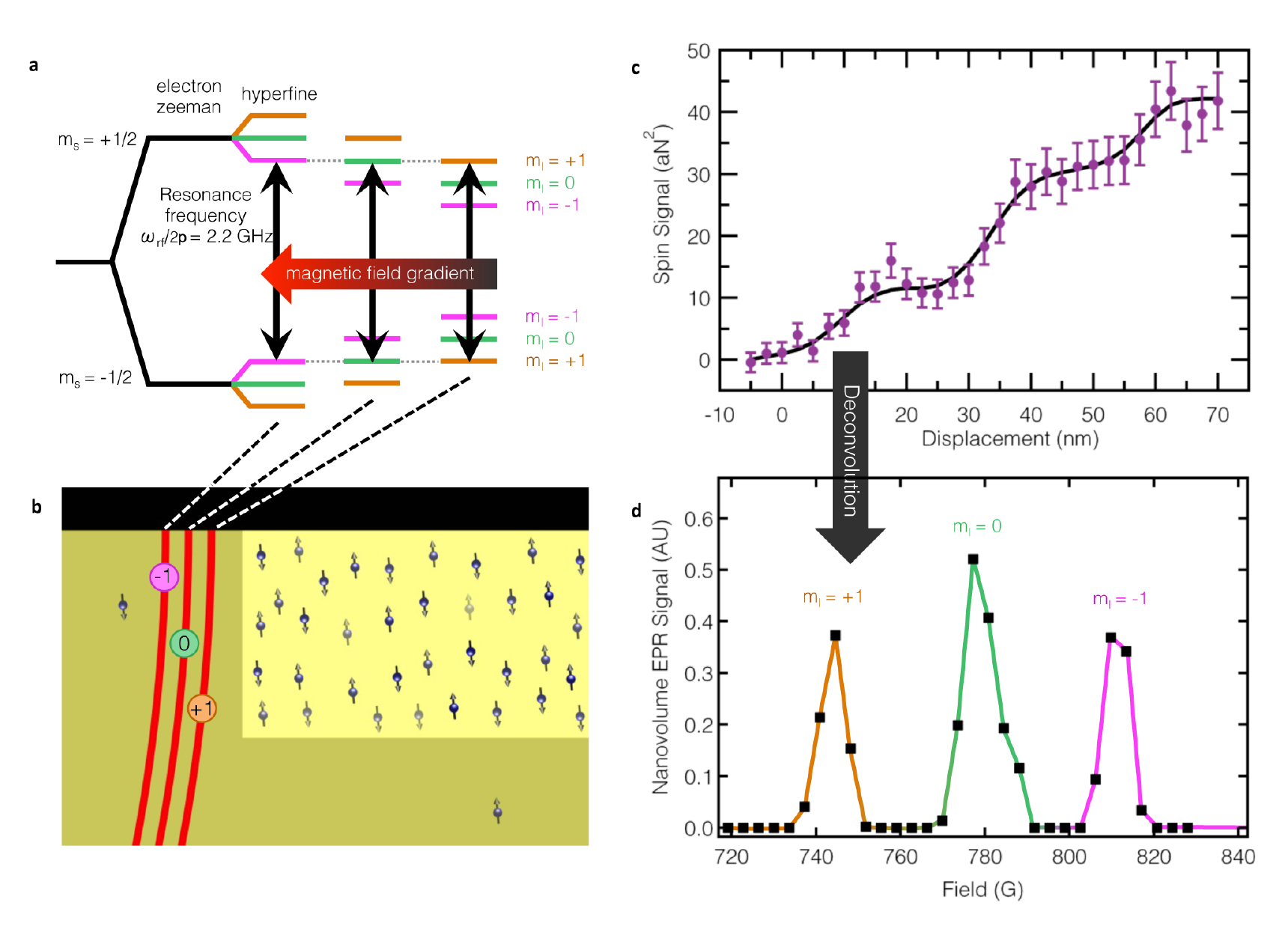}
\par\end{centering}

\caption{\textbf{Hyperfine spectrum measurement for spin wire}. \textbf{a},
Energy level diagrams for three hyperfine-split spin populations,
which are simultaneously on resonance, yet spatially separated by
the gradient. \textbf{b}, Schematic of the three volumes which are
walked into the spin wire. \textbf{c}, The spin signal increases in
a step-like fashion as each volume enters the wire. \textbf{d}, A
nanovolume EPR spectrum can be obtained by deconvolving the spin signal
with the known force sensitivity (see SI) and the three hyperfine
peaks are resolved.}
\end{figure*}
Once the detection volume enters the spin wire, the signal grows with
the number of detected spins, eventually reaching a plateau when completely
within the wire. The correlation time shows a complex behaviour as
the detection volume moves into the wire. These results can be understood
within the framework of flip-flop-mediated spin transport: ensemble
correlation is lost as spins diffuse into or out of the detection
volume. Just inside a displacement of 0 nm (Fig. 1b), the spins inside
the volume easily and rapidly interact with nearby outside neighbours,
resulting in a relatively short correlation time. With increasing
overlap of the detection volume and the spin wire, spins must diffuse
further to exit the detection volume and change the overall magnetization
of the ensemble, thus increasing $\tau_{{\rm m}}$. The correlation
time peaks when approximately $68\%$ of the detection volume has
entered the spin wire (Fig. 1c). This is attributed to the force sensitivity
profile of the detection volume: spins in the middle of the volume
are most sensitively detected (solid red indicates highest sensitivity
in Fig. 1), while the edges are least sensitive (see SI). As the detection
volume approaches $100\%$ overlap (Fig. 1d), spins are able to diffuse
out either side of the most sensitive region, reducing the correlation
time by roughly a factor of $2$. Scanning deeper into the spin wire
results in no further change, since the measurement geometry becomes
translationally invariant (Fig. 1e). 

A global measurement of the spin polarization, encompassing the entire
spin wire, would not uncover this magnetization-preserving spin transport
process. In the opposite limit of measuring a single spin, $T_{1}$-type
spin relaxation becomes indistinguishable from flip-flop induced polarization
changes. By systematically varying the size of the overlap of the
detection volume with the spin wire, and hence the number of detected
spins, our measurement of $\tau_{{\rm m}}$ reveals the observed signature
of spin transport. This is similar to measurements of spin lifetime
by electrically- or optically-detected Hanle signals\cite{lou2007electrical,huang2010time,furis2007local},
where diffusion of spins away from the detection volume (electrical
contact or optical probe spot) can influence spin dephasing.

Since the average polarization gradient vanishes at thermal equilibrium,
the conventional model of diffusion driven by polarization gradient
predicts that $\tau_{{\rm m}}$ will be independent of detection volume
position within our thermal equilibrium wire. Such a model neglects
the dominant role spin fluctuations play in nanoscale ensembles. In
order to fit the measured data in Fig. 1f, we use a Monte Carlo simulation
which models the flip-flops between individual spins as a Markov process
(see SI). From this fit we can extract the flip-flop time, $T_{{\rm ff}}=0.21$
ms, and the corresponding diffusion constant of $D=\frac{a^{2}}{T_{{\rm ff}}}=4.6\times10^{-9}\frac{{\rm cm}^{2}}{{\rm s}}$,
where $a=n^{-1/3}=9.82$ nm is the average nearest-neighbour separation
of spins with a density $n=1.06\times10^{18}$ cm$^{-3}$ (6 ppm).
We find good agreement comparing this to the theoretical flip-flop
time\cite{abragam} $T_{{\rm ff,th}}=30\, T_{2}=0.36$ ms, where we
have used T$_{2}=11.9$ $\mu$s, which arises from dipolar interactions
(at a density of 6 ppm) and has been experimentally verified\cite{van1997dependences}.
We furthermore find a diffusion length $L=\sqrt{DT_{1}}=720$ nm,
which is significantly larger than the lateral dimensions of the wire
(depth $250$ nm, width $200$ nm, and length $4$ $\mu$m). This
diffusion length is competitive with metallic spin transport devices\cite{jedema2002electrical}. 

The above expression for $T_{{\rm ff,th}}$ assumes zero magnetic
field gradient for the system, but in our experiment we have a strong
gradient ($\sim1.3$ G/nm). In the presence of a field gradient, neighbouring
spins experience a field difference $\Delta B$, and thus different
Zeeman energy splittings. This can suppress flip-flops because they
no longer conserve energy\cite{r:budakian:037205,PhysRevB.12.78,tyryshkin2011electron}.
However, inhomogeneous line broadening, if on the order of $\Delta B$,
can make up this energy difference and allow flip-flops to proceed.
Since the measured flip-flop time $T_{{\rm ff}}$ agrees closely to
the expected zero-gradient value $T_{{\rm ff,th}}$, we expect a spectral
linewidth $\Delta B\gtrsim6.5$ G.

To measure the EPR spectrum of the spin wire, we implemented a modified
iOSCAR protocol which improves spatial resolution and therefore provides
the capability to resolve spectral features (since spatial and spectral
resolution are directly related in magnetic resonance imaging). A
conventional iOSCAR measurement utilizes the entire detection volume
swept out by the cantilever oscillation, as this couples to the largest
number of spins and creates the largest force signal. By truncating
the detection volume to the most-sensitive portion, we couple only
to spins in this smaller region while maintaining high force sensitivity.
Utilizing this technique (further details in SI), which we call partially-interrupted
OSCAR (piOSCAR), we again scan the detection volume into the spin
wire. The improved resolution enables us to resolve a staircase structure
with three steps (Fig. 2c), corresponding to the three peaks of the
P1 center\textquoteright{}s hyperfine spectrum as discussed below.

The P1 centre is a substitutional nitrogen defect with an unpaired
spin-$\frac{1}{2}$ electron hyperfine coupled to the spin-$1$ nitrogen
nucleus. The hyperfine interaction results in a triplet of peaks in
the EPR spectrum (Fig. 2a) corresponding to the three nuclear spin
states, $m_{{\rm I}}=-1,\,0,\,1$. The hyperfine splitting for our
diamond crystal orientation relative to the external field is 33 G
(see SI and ref. \cite{cook1966electron}). In the magnetic field
gradient of our probe magnet, this splitting results in three spatially
distinct spin detection volumes, with each volume defined by the contour
of magnetic field which satisfies the resonance condition for a particular
hyperfine transition (Fig. 2b). We observe a step-like increase in
the spin signal as each volume enters the wire and the cantilever
becomes coupled to an additional hyperfine transition (Fig. 2c). The
spacing between steps (s = 25 nm) provides an accurate means of measuring
the probe field gradient, because the spacing is set by the ratio
of the known hyperfine splitting ($33$ G) to the gradient, which
we find to be $1.3$ $\frac{\mbox{G}}{\mbox{nm}}$. This is consistent
with our estimations of gradient obtained using a standard technique\cite{stipe2001electron}.
The staircase structure in figure 2c is a convolution of the implanted
spin density (approximately a step function), our force sensitivity
profile, and the EPR spectrum of spins in the sample; deconvolution
reveals the EPR spectrum shown in Fig. 2d (details in SI). From the
spectrum we find an average linewidth of $7.6$ G, which agrees with
the expected linewidth from above, and corroborates the transport
we measure in the spin wire. 

In conclusion, we have measured flip-flop-mediated spin transport
in an insulating diamond spin wire in the complete absence of charge
transport. This transport arises from the intrinsic spin dynamics
of dipole-coupled P1 centres at thermal equilibrium, which we observe
by measuring spin noise without imposing a polarization gradient.
By spatially resolving the three electron spin populations (corresponding
to the three spin states of the hyperfine-coupled nitrogen nucleus)
we obtained the EPR spectrum of less than $100$ net spins in the
spin wire. The resulting linewidth confirms that flip-flop mediated
spin diffusion is responsible for the observed spin transport. The
spectrum also enables accurate measurement of the applied magnetic
field gradient. These measurements provide insight into the mechanisms
of spin transport relevant for the development of nanoscale spin device
elements. 

\textbf{Methods} The ultra-soft silicon cantilever has a spring constant
of $100$ $\mu$N/m, resonance frequency of about $3$ kHz, and approximate
dimensions of $90$ microns in length, $1$ micron width, and $100$
nm thickness. Displacement of the cantilever is measured through a
$1550$ nm laser interferometer. The MRFM measurements were taken
at temperatures of $4.2$ K. The measured moment of the SmCo$_{5}$
magnetic particle was $3.6\times10^{-12}$ J/T. The niobium resonator
coil had about $2.5$ turns and a $300$ micron diameter, and the
resonance bowl had a diameter of about $3$ microns. The diamond sample
studied in this experiment had a background nitrogen concentration
of $0.3$ ppm ($5.27\times10^{16}$ cm$^{-3}$). To create the high
spin-density wire, the sample was first prepared using electron beam
patterning and then exposed to nitrogen ion implantation at a variety
of energies to create a uniform spin density. Finally the sample was
annealed to yield an approximately uniform, channel ($6$ ppm or $1.06\times10^{18}$
cm$^{-3}$) with width, depth, and length of $200$ nm, $250$ nm,
and $4$ microns, respectively. The sample was purchased commercially
from Element Six, grown via chemical vapor deposition, and the nitrogen
ion implantation was performed by Leonard Kroko Inc. 

\textbf{Acknowledgements} The research presented in this manuscript
was financially supported by the Army Research Office (grant number W911NF-09-1-0147),
the National Science Foundation (grant number DMR-0807093), and the Center
for Emergent Materials (CEM), an NSF funded MRSEC (grant number DMR-0820414). Technical 
support was provided by the NanoSystems Laboratory at the Ohio State University. The authors
would like to thank R. J. Furnshahl for valuable discussions related
to the simulations. 

\bibliographystyle{naturemag}

\end{document}